\documentclass[
amsmath,
prd,nofootinbib,floatfix,11pt
]{revtex4}

\newcommand\beq{\begin{eqnarray}}
\newcommand\eeq{\end{eqnarray}}

\def\lsim{\mathrel{\rlap{\lower4pt\hbox{$\sim$}}
    \raise1pt\hbox{$<$}}}                
\def\gsim{\mathrel{\rlap{\lower4pt\hbox{$\sim$}}
    \raise1pt\hbox{$>$}}}            

\allowdisplaybreaks
\usepackage{graphicx}
\usepackage{setspace}
\usepackage{placeins}
\usepackage{hyperref}
\usepackage[T1]{fontenc} 
\usepackage{float}
\usepackage{subfig}
\usepackage{placeins}
\usepackage{latexsym}
\usepackage{color}
\definecolor{cblue}{RGB}{100,5,255}
\definecolor{cred}{RGB}{255,50,40} 
\definecolor{cgreen}{RGB}{40,255,40} 
     
\usepackage{dcolumn}
\usepackage{bm}

\begin{document}

\renewcommand{\theequation}{\arabic{section}.\arabic{equation}}
\renewcommand{\thefigure}{\arabic{section}.\arabic{figure}}
\renewcommand{\thetable}{\arabic{section}.\arabic{table}}

\title{\Large \baselineskip=10pt 
Study in the noncanonical domain of Goldstone inflation}

\author{Sukannya Bhattacharya$^{1,2,3}$ and Mayukh~R.~Gangopadhyay$^{1,4}$}
\affiliation{
{\it $^1$Theory Division, Saha Institute of Nuclear Physics, Bidhannagar, 700064, India.} \\
{\it $^2$Homi Bhabha National Institute, Training School Complex, Anushaktinagar, Mumbai
400094, India.}\\
{\it $^3$Theoretical Physics Division, Physical Research Laboratory, Navrangpura, Ahmedabad - 380009, India}\\
{\it$^4$Centre For Theoretical Physics, Jamia Millia Islamia, New Delhi-110025, India.}
}


\begin{abstract}\normalsize \baselineskip=10pt 
Recent observations of the cosmic microwave background (CMB) indicate that a successful theory of cosmological inflation needs to have flat potential of the inflaton scalar field. Realizing the inflaton to be a pseudo-Nambu Goldstone boson (pNGB) could ensure the flatness and the sub-Planckian scales related to the dynamics of the paradigm. In this work, we have taken the most general form of such a scenario: Goldstone inflation proposed in \cite{coon} and studied the model in the noncanonical domain. Natural inflation is a limiting case of this model, which is also studied here in the noncanonical regime. Our result is compared with the recent release by Planck collaboration and it is shown that for some combination of the model parameters, a Goldstone inflationary model in the noncanonical realisation obeys the current observational bounds. Then, we studied the era of reheating after the end of inflation. For different choice of model parameters, constraints on the reheating temperature ($T_{\rm re}$) and  number of e-folds during reheating($N_{\rm re}$) for the allowed inflationary observables (e.g.~scalar spectral index($n_s$) and tensor to scalar ratio($r$)) are predicted for this model.
\end{abstract}

\maketitle

\vspace{0.0001in}

\tableofcontents

\baselineskip=15.4pt

\section{Introduction}
\setcounter{equation}{0}
\setcounter{figure}{0}
\setcounter{table}{0}
\setcounter{footnote}{1}

Inflationary paradigm offers a very attractive solution to resolve the hot big bang cosmology puzzles (for a review reader is advised to consult these books: \cite{Liddle}, \cite{cmbinflate}). From the time of its first proposal almost 40 years ago \cite{guth}, the idea has been one of the prime focus of research in theoretical particle physics and cosmology (see \cite{linde}-\cite{mukha81} for important early work). With the stupendous advancement in the observational cosmology, now we can constrain different inflationary models in detail from real data. One of the problems with the standard inflationary models is that most of the textbook models are ruled out or disfavored by the recent observations of CMB such as {\it Planck} and {\it WMAP} \cite{PlanckXX, WMAP9}. In early 1990s one elegant solution was proposed by Freese \textit{et. al.}~\cite{freese} from the idea of symmetry breaking to produce the inflation potential where the inflaton is a Goldstone boson (Natural inflation). Due to the shift symmetry property embedded through the symmetry breaking, the flatness of the potential is maintained, which is essential for the model building of inflation. But after the recent data release by Planck collaboration \cite {planck18}, natural inflation is almost ruled out in the standard $\Lambda$CDM model. The BIC (Bayesian information criterion) calculated for such models puts it right on the fence for getting invalidated by data.

 Natural inflation is one particular limiting case of a general class of inflation models known as Goldstone inflation. To have a successful Goldstone inflation, all scales related to the theory have to be sub-Planckian, thus keeping the inflaton guarded against the UV correction from the quantum gravity effects. Now in standard scenario of inflation, the scalar field is taken to be canonical. But, it was realized after the initial proposal of kinetic driven noncanonical inflation (NCI) in 1999 by Garriga and Mukhanov \cite{mukhanov} that NCI's are more natural to fit with the fundamental theories like string theory. 

After the proposal of tachyon inflation in \cite{Sen}, noncanonical realization of different inflationary models have gained growing interest. Thus, in this work we move on to study the Goldstone inflation in the noncanonical  domain and check the viability of the model from direct constraints by the current observation. In this work we tried to explore the general Goldstone inflation in noncanonical domain and then studied noncanonical natural inflation as a special case. 

The rest of the paper is organized as follows. In~\ref{sec:review} we will make a brief review of the standard Goldstone inflation along with the basic ingredients to build up the noncanonical inflationary dynamics. In \ref{analysis}, we have reported the analysis part and in~\ref{results} we present the main results obtained through that analysis, after that a brief discussion on the reheating is carried out in the~\ref{reheat}. Finally the conclusions are drawn in the final section.
\label{sec:intro}

\section{Revisiting the canonical Goldstone and noncanonical inflation}
\setcounter{equation}{0}
\setcounter{figure}{0}
\setcounter{table}{0}
\setcounter{footnote}{1}
\label{sec:review}
\subsection{Reviewing Goldstone inflation}

The originally proposed model of natural inflation has an axion as the inflaton, which is the Goldstone of a spontaneously broken Peccei-Quinn symmetry. But, as mentioned in the previous section, it is almost ruled out in the standard $\Lambda$CDM paradigm by recent CMB observations. The model is still in the 2$\sigma$ allowed region with an associated breaking scale of $~10M_{Pl}$ or higher. This is problematic as the effective field theory dynamics could get completely jeopardized by the effects of the quantum gravity (QG) which should robustly kick in to the picture in the super-Planckian regime. QG in general does not conserve global symmetry and therefore to have a super-Planckian breaking scale in case of a vanilla natural inflation model is philosophically very disturbing.

Different exquisite models have been proposed to explain the super-Planckian breaking scale, such as extranatural inflation \cite{nima}, hybrid axion models \cite{lindehyb, kim}, N-flation (\cite{nflation1} -\cite{nflation3}), axion monodromy \cite{axionmono} and other pseudonatural inflation models in supersymmetry \cite{pngb}. Some or most of these models require a large amount of tuning or the existence of extra dimensions. But even with these theoretical explanations, with the recent release of Planck data, the idea of natural inflation faces a survival crisis. The vanilla model is disfavored by the Planck 2018 plus BK14 data with a Bayes factor $ln B = -4.2$ (Models are strongly disfavoured when $ln B < -5$). Therefore, it is high time to reevaluate the original motivation and development of the models of Natural inflation where the potential is generated through the breaking of a global symmetry.

In \cite{coon}, there is a proposal of a model where a generalized Goldstone inflation is developed from the idea of minimal composite Higgs model \cite{csaki, contino}.

The form of the potential to give a successful inflation is given as :
\begin{equation}
V(\phi) = \Lambda^4 (C_{\Lambda} +\alpha \cos(\phi /f) +\beta \sin^2(\phi/f))
\label{eq1}
\end{equation}
In \cite{coon}, it has been shown that with an appropriate amount of fine tuning, one obtains a successful model of Goldstone inflation with a sub-Planckian breaking scale related to the global symmetry breaking. Now, with the recent results from Planck 2018, even a canonical Goldstone inflation faces problem to survive. Since, this model is motivated by the minimal composite Higgs model, it is expected to have noncanonical origin in the dynamics of the inflation. It is also clear from~\ref{eq1} that, for the choice of the parameter $\alpha= 1, \beta= 0$ one gets back the standard form of the natural inflation potential. Here, we note that this work is a phenomenological approach toward kinetic features of Goldstone inflation and a description of the complete high energy theory that predicts the pNGB with a modified kinetic term is beyond the scope of this work.
\subsection{Revisiting NCI}
Here, we will briefly review the noncanonical inflation before introducing the Goldstone inflation in the noncanonical regime.
NCI model features a single scalar field with the action \cite{mukhanov,picon,li}:
\begin{equation}
S = \int \sqrt{-g}~p(\phi, X)d^4 x~,
\label{eq2}
\end{equation} 
where $\phi$ is the inflaton field. Here $p(\phi, X) = K(X, \phi)- V(\phi)$, where $V(\phi)$ is the potential and $X\equiv \frac{1}{2}\partial_{\mu}\phi\partial^{\mu}\phi$. Now, it is very import to understand that the kinetic term $K(X, \phi)$ can be any arbitrary function of $X$ and $\phi$ with proper dimensional attributions to the prefactors. Here, let us write $K(X,\phi)$ as :
\begin{equation}
K(X,\phi)= K_{nc}(\phi)K_{kin}(X)~,
\label{eq3}
\end{equation} 
here, $K_{nc}(\phi)$ can be any arbitrary function of $\phi$. On the other hand, assuming a power law function, $K_{kin}(X)\equiv K_{n+1} X^n$, where $n$ is the power. Thus for $n>1$, we find higher order contribution of the pure kinetic term even with dimensionful constant $K_{n+1}=1$. From Eq.~\eqref{eq3}, it is expected to get back the canonical picture once we set $n=1, K_{n+1}=1$ and $K_{nc}(\phi)=1$ respectively. 

For the purpose of this paper we separate the contributions of the field dependent kinetic term $K_{nc}(\phi)$ and of the derivative dependent kinetic term $K_{kin}(X)$. The scenario with $K_{nc}(\phi)$ switched on and $K_{kin}(X)=X$ is termed as {\it Case 1}. The case where we consider $K_{nc}(\phi)= 1$ and $K_{kin}(X)$ is non trivially switched on is called {\it Case 2}.
 \subsubsection{{\emph Case 1}}
 \label{revisitcase1}
In this case, $K_{nc}(\phi)$ is switched on and $K_{kin}(X)\equiv X$. Thus, in this case there is no higher order kinetic term present and the effective Lagrangian for generic $K_{nc}(\phi)$ and $V(\phi)$ can be written as:
\begin{equation}
\mathcal{L}= K_{nc}(\phi)X - V(\phi).\label{Lnoncangen}
\end{equation}
Then the equation of motion (EoM) for the field $\phi$ turns out to be
\begin{equation}
\ddot{\phi}+3H\dot{\phi}+\frac{K_{nc,\phi}}{2K_{nc}}\dot{\phi}^2+\frac{V_{,\phi}}{K_{nc}}=0,\label{EoMnoncangen}
\end{equation}
where $V_{,\phi} = dV/d\phi$ and $K_{nc,\phi} = dK_{nc}/d\phi$. If the canonical field is given as $\psi$ such that $\frac{1}{2}\partial_{\mu}\psi\partial^{\mu}\psi = \frac{1}{2}K_{nc}(\phi)\partial_{\mu}\phi\partial^{\mu}\phi$ then the slow roll parameters are modified as:
\begin{eqnarray}
\epsilon_V &=& \frac{M_{Pl}^2}{2}\bigg( \frac{V_{,\psi}}{V}\bigg)^2 = \frac{M_{Pl}^2}{2K_{nc}}\bigg( \frac{V_{,\phi}}{V}\bigg)^2, \label{SR1noncangen}\\
\eta_V &=& M_{Pl}^2\bigg( \frac{V_{,\psi\psi}}{V}\bigg) = \frac{M_{Pl}^2}{V}\bigg(\frac{V_{,\phi\phi}}{K_{nc}} - \frac{V_{,\phi}K_{nc,\phi}}{2K_{nc}^2}\bigg).\label{SR2noncangen}
\end{eqnarray}
The number of inflationary e-folds in the slow roll regime is
\begin{equation}
N=\frac{1}{M_{Pl}}\int_{\phi_i}^{\phi_e} \frac{V}{V_{,\phi}\sqrt{K_{nc}}}d\phi. \label{Nnoncangen}
\end{equation}
The above relations (Eq.~\ref{Lnoncangen} to Eq.~\ref{Nnoncangen}) are true for any inflaton potential $V(\phi)$ and we will speculate the particular form of Goldstone inflation in this noncanonical setting in Sec.~\ref{GI-NCcase1-analysis}. The inflationary observables in this case are
\begin{align}
n_{s} - 1 & = 2\eta_V - 6\epsilon_V \label{case1ns}\\
r & = 16\epsilon_V\label{case1r}
\end{align}
\subsubsection{{\emph Case 2}}
\label{review_case2}
In this case, $K_{nc}(\phi)\equiv 1$ and $K_{kin}(X)\equiv K_{n+1} X^n$ (for a comprehensive review reader is suggested to consult \cite{li}). Here, the total background dynamics can be constructed in terms of $p(\phi, X) = K(X)- V(\phi)$. The Hubble equation is given as:
\begin{equation}
H^2 = \rho /3,
\label{kHubble}
\end{equation}
where
\begin{equation}
\rho = 2Xp_{,X}-p.
\label{krho}
\end{equation}
The speed of sound is 
\begin{equation}
c_s^2 = \frac{p_{,X}}{\rho_{,X}}=\frac{K_{,X}}{2XK_{,XX}+K_{,X}},
\label{ksoundspeed}
\end{equation}
using Eq.~\eqref{krho}. For the given form of $K(X)$, the sound speed is a constant $c_s^2=1/(2n-1)$ and the equation of motion (EoM) for the inflaton in this case is modified to:
\begin{equation}
\ddot{\phi}+\frac{3H}{2n-1}\dot{\phi}+\frac{V_{,\phi}}{(2n^2-n)K_{n+1}X^{n-1}}=0.
\label{kEOM}
\end{equation}
Now, the slow roll parameters are needed to be calculated to get the expressions for the observables.
The two potential slow roll parameters are given as:
\begin{eqnarray}
\epsilon_{V} &=&\frac{1}{2} \gamma(n) \left(\frac{{V_{,\phi}}^{2 n}}{V^{(3 n-1)}}\right)^{\frac{1}{2 n-1}}\\
\eta_{V} &=&\gamma(n)\left(\frac{V_{,\phi\phi}^{(2n-1)}}{V^n V_{,\phi}^{(2n-2)}}\right)^{\frac{1}{2 n-1}},
\end{eqnarray}
where $\gamma(n)=\bigg(\frac{6^{n-1}}{nK_{n+1}}M_{\rm Pl}^{2n}\bigg)^{\frac{1}{2n-1}}$.

The scalar and tensor power spectra are given as:
\begin{eqnarray}
\mathcal{P}_s &=& \frac{1}{8\pi ^2M_{Pl}^2}\frac{H^2}{\epsilon_{V} c_s}\vert _{c_sk=aH},\label{kscalP}\\
\mathcal{P}_t &=& \frac{2}{\pi ^2 M_{Pl}^2}H^2\vert _{k=aH}\label{ktensP}.
\end{eqnarray}
Then the inflationary observables can be calculated to be:
\begin{eqnarray}
n_s -1&=& \frac{1}{2 n-1}[2n \eta_V- 2(5 n-2) \epsilon_V] \label{ns_kine}\\
r &=& 16 c_s \epsilon_V\label{rkine}
\end{eqnarray}
Finally, the number of e-folds can be expressed in this case as:
\begin{equation}
N= \int^{\phi_i}_{\phi_e} \frac{1}{\gamma(n)} \sqrt{V} \left(\frac{\sqrt{V}}{V_{,\phi}}\right)^{\frac{1}{2 n-1}} \, d\phi
\end{equation}
Here, $\phi_i$ and $\phi_e$ represents the field values of the inflaton field at the horizon exit and end of inflation respectively.

\section{Analysis for Goldstone inflation}
\label{analysis}
\setcounter{equation}{0}
\setcounter{figure}{0}
\setcounter{table}{0}
\setcounter{footnote}{1}

Here, we analyze the effect of non-canonial scenarios {\it Case 1} and {\it Case 2} on the dynamics of Goldstone inflation. We consider the potential for the Goldstone inflation in the form of Eq.~\eqref{eq1} with $C_\Lambda = \alpha = 1$, $\beta \equiv \frac{\beta}{\alpha}$. 
\subsection{{\emph Case 1}}
\label{GI-NCcase1-analysis}
\begin{figure}[h!]
\centering{
 \includegraphics[width=0.65\textwidth]{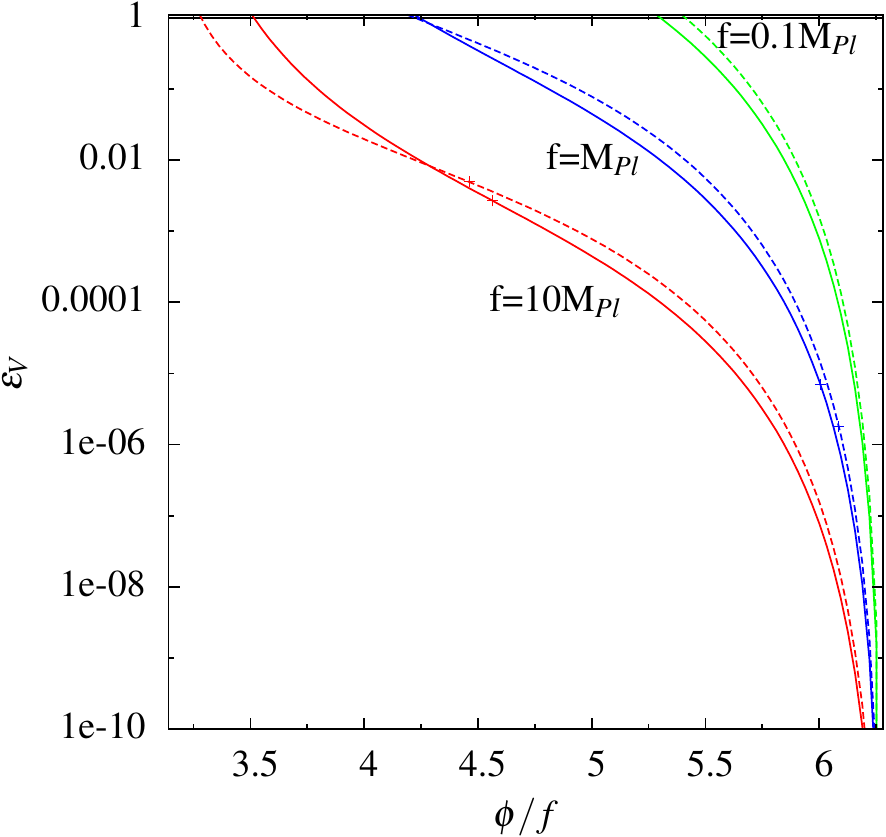} }
 \caption{\small{The variation of $\epsilon_V$ as a function of the field. The dashed lines represent canonical Goldstone inflation whereas the solid lines represent noncanonical Goldstone inflation with $K_{nc}$ switched on and $K_{kin}= X$ ({\it Case 1}). The pivot field values for $55$ e-folds of inflation for the cases $f=M_{Pl}$ and $f=10M_{Pl}$ are marked with crosses in the curves.}}
\label{plot_epscase1} 
\end{figure}
 Using the following noncanonical form:
\begin{eqnarray}
~K_{kin}(X) &=& X, \\
K_{nc}(\phi) &=& 1 +\alpha \cos(\phi /f) +\beta \sin^2(\phi/f)= \frac{V(\phi)}{\Lambda^4}, \label{ncform}
\end{eqnarray}

we arrive at the EoM:
\begin{equation}
\ddot{\phi}+3H\dot{\phi}+\bigg(\frac{\dot{\phi}^2}{2}+\Lambda^4\bigg)\bigg[\frac{-\alpha \sin(\phi/f)+\beta \sin(2\phi/f)}{f(1 +\alpha \cos(\phi /f) +\beta \sin^2(\phi/f))}\bigg] = 0.\label{EoMncgi}
\end{equation}
Here, the reason behind the particular choice of $K_{nc}(\phi) = V(\phi)/\Lambda^4$ is that inflation is an effective field theory (EFT) and is valid up to the breaking scale $f$. The physics above $f$ is integrated out and the resulting higher dimensional operators are neglected and if one takes care of that, there will be presence of noncanonical kinetic terms. In our case the choice $K_{nc}(\phi)$ is from the phenomenological perspective to take the advantage of the periodicity in the kinetic part.

The slow roll parameters are:
\begin{eqnarray}
\epsilon_V &=& \frac{\Lambda^4M_{Pl}^2}{2}\bigg(\frac{(-\alpha \sin(\phi/f)+\beta \sin(2\phi/f))^2}{f^2(1 +\alpha \cos(\phi /f) +\beta \sin^2(\phi/f))^3}\bigg), \label{epsv_ncgi}\\
\eta_V &=& M_{Pl}^2\Lambda^2 \frac{-\alpha \cos(\phi/f)+2\beta \cos(2\phi/f)-\alpha^2-\beta^2(1- \cos(2\phi/f)+\alpha\beta \cos(\phi/f)(1+ \cos^2(\phi/f))}{f^2(1 +\alpha \cos(\phi /f) +\beta \sin^2(\phi/f))^3}.\nonumber \\
\label{etav_ncgi}
\end{eqnarray}
It is worth mentioning that the typical prescription here and in Sec.~\ref{revisitcase1} to express the dynamical quantities of inflation in terms of a canonical field variable is a common practice for any such noncanonical modifications. In terms of the canonical variable, the form of the potential $V(\psi)$ is evidently different from $V(\phi)$. In our case, this scenario aids to obtain an effective slower roll for the inflaton $\phi$, such that $\epsilon_V$ is lower than the default canonical case Fig.~\ref{plot_epscase1} for the relevant field excursion.

In Fig.~\ref{plot_epscase1}, the variation of $\epsilon_V$ is shown as a function of the normalized field value $\phi/f$. For the breaking scale $f=10M_{Pl}$ (red line), the $\epsilon_V$ at pivot (marked with cross) is lower in case of noncanonical Goldstone inflation than the canonical case, therefore pointing toward a lower energy scale of inflation. But for $f\leq M_{Pl}$, the pivot energy scale for noncanonical case is higher than the canonical case. This has been depicted by the blue and green lines in the Fig.~\ref{plot_epscase1}. It should be noted that varying $\alpha$ also as a parameter may improve the predictions for observables.

\subsection{{\emph Case 2}}
\begin{figure}[h!]
\centering{
 \includegraphics[width=0.6\textwidth]{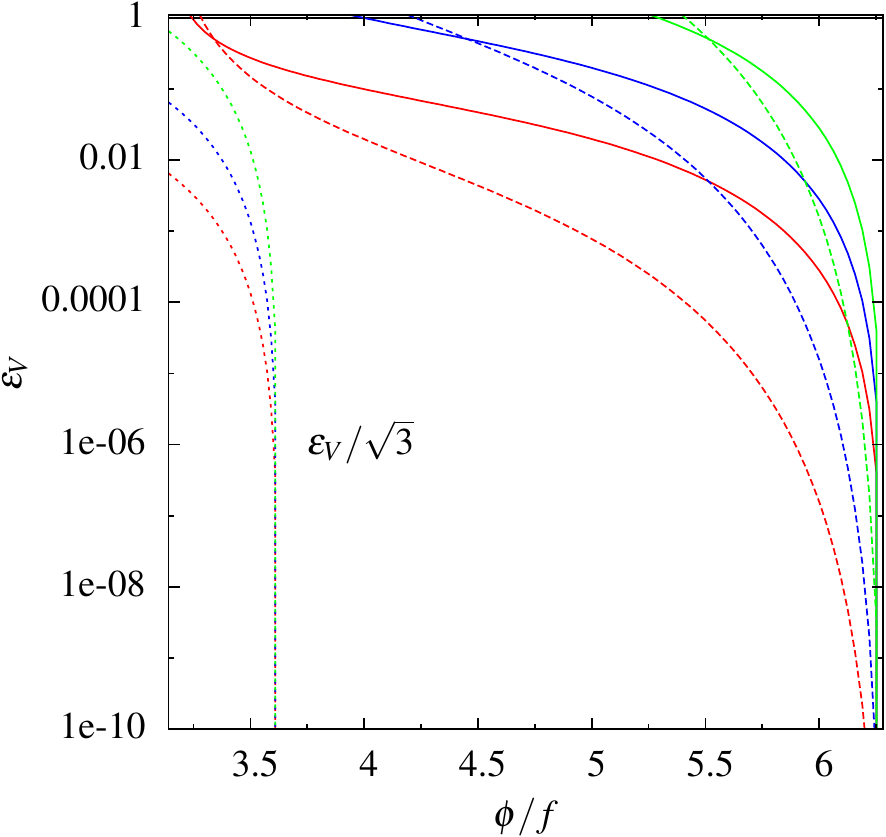} }
 \caption{\small{The variation of $\epsilon_V$ as a function of the field. The dashed lines represent canonical Goldstone inflation whereas the solid lines represent noncanonical Goldstone inflation with $K_{kin}$ switched on and $K_{nc}=1$. The plots are for $f=10M_{Pl}$ (in red), $f=M_{Pl}$ (in blue) and $f=0.1M_{Pl}$ (in green).}}
\label{plot_epscase2} 
\end{figure}
The natural inflation is a limiting case of the generalized goldstone inflation with $\alpha =1$ and $\beta = 0$. We start with the analysis of natural inflation to clarify the dependence of the inflationary observables on the parameters of the model, which is also applicable by extension to generic Goldstone inflation. For natural inflation, the potential and the kinetic functions are given as:
\begin{align}
V &= \Lambda^4 (1+ \cos(\phi/f))\\
K &= K_{n+1}X^n
\end{align}
Then, the slow roll parameters are:
\begin{eqnarray}
\epsilon_V &=& \frac{1}{2} \frac{V_{\phi}}{V} \gamma(n) \bigg(\frac{V_{\phi}}{V^n}\bigg)^{\frac{1}{2n-1}}\nonumber \\
                  &=& \frac{1}{2}\frac{1}{(K_3\Lambda^4)^{1/3}}\bigg[\frac{\sin^4(\phi/f)}{(1+ \cos(\phi/f))^5}\bigg]^{1/3}
\end{eqnarray}
\begin{eqnarray}
\eta_V &=& \gamma(n)\bigg( \frac{V_{\phi \phi}^{(2n-1)}}{V^nV_{\phi}^{(2n-1)}}\bigg)^{\frac{1}{2n-1}}\nonumber \\
                  &=&\frac{1}{(K_3\Lambda^4)^{1/3}}\bigg[\frac{\cos(\phi/f)}{\sin(\phi/f)(1+ \cos(\phi/f))^{2/3}}\bigg],
\end{eqnarray}
where in each of the above two equations, the second line is the expression for $n=2$.
The ratio of the scalar power spectra in case of kinetic natural inflation ($n=2$) to the canonical natural inflation ($n=1$) can be written as:
\begin{eqnarray}
\frac{P_{s}^{n=2}}{P_{s}^{n=1}} &=& \frac{1}{\gamma(2)c_s}\times \frac{V^{8/3}}{V_{\phi}^{4/3}}\times \frac{V_{\phi}^2}{V^3}=   \frac{1}{\gamma(2)c_s}\times \frac{V_{\phi}^{2/3}}{V^{1/3}} = \bigg(\frac{K_3}{3}\bigg)^{1/3}\times 3^{1/3}
\times\frac{V_{\phi}^{2/3}}{V^{1/3}}\nonumber \\
&=& 3^{1/6} \times K_{3}^{1/3}\bigg[\frac{(\Lambda^{4}/f)^2 \sin^2(\phi/f)}{\Lambda^4 (1 + \cos(\phi/f))}\bigg]^{1/3}
\nonumber \\
&=& 3^{1/6} \times \bigg(\frac{1}{f^{2/3}}\bigg)\times (K_3\Lambda^4)^{1/3}\times\bigg[\frac{\sin^2(\phi/f)}{1+\cos(\phi/f)}\bigg]^{1/3}\nonumber \\
\label{eq31}
\end{eqnarray}
Thus, from the Eq.\eqref{eq31} it is clear that:

\begin{equation}
\frac{P_{s}^{n=2}}{P_{s}^{n=1}}\propto \frac{(K_3\Lambda^4)^{1/3}}{f^{2/3}}
\end{equation}
From the dependences of $\epsilon_V$ and $\eta_V$ for natural inflation here, on the factor $(K_3\Lambda^4)^{1/3}$ and on $f$, it is evident that the slow roll parameters have values ($>1$) not compatible with the slow roll condition for most of the inflaton's journey on the slope of the potential. To summarize the  point, let us take $f = 10 M_{Pl}$. For the sake of a simplistic analysis, let us assume the term in the square bracket in the Eq.\eqref{eq31} is $\mathcal{O}(1)$. Then, the first slow roll parameter is:
$\epsilon_V \simeq \frac{3^{1/3}}{2}\times10^{-4}\times\frac{1}{(K_3\Lambda^4)^{1/3}}.$
Thus, for $K_3 =1$, for any realistic scale of inflation (value of $\Lambda^4$) the pivot value of $\epsilon_V$ is quite large to have $50- 60$ e-folds of inflation, which is required from observations. Therefore, it is difficult to achieve enough number of inflationary e-folds for {\it Case 2} of kinetic natural inflation.

By analogy, for the case of Goldstone inflation in the noncanonical regime of {\it Case 2}, the combination $(K_3\alpha\Lambda^4)^{1/3}$ influences the dynamics of inflation in a similar way. The factor $\alpha^{1/3}$ appears here since we have considered it as an overall factor in the potential in Eq.~\eqref{eq1} and varied the normalized value of $\beta \equiv \beta/\alpha$. The variations of the first slow roll parameter $\epsilon_V$ as a function of $\phi/f$ is shown in Fig.~\ref{plot_epscase2}, where, unlike {\it Case 1}, $\epsilon_V$ for a noncanonical {\it Case 2} is higher than that for a canonical case for a particular value of $\phi/f$. But, as shown in Eq.~\eqref{rkine} in Sec.~\ref{review_case2}, the energy scale of inflation depends on the speed of sound $c_s=\frac{1}{\sqrt{2n-1}}=1/\sqrt{3}$. This factor appears in the EoM Eq.~\ref{kEOM} and also in the expression of the pivot quantities. A variation of $c_s\epsilon_V=\epsilon_V/\sqrt{3}$ in Fig.~\ref{plot_epscase2} shows that for a particular field value, the noncanonical Goldstone inflation ({\it Case 2}) points to a lower effective energy scale of inflation (plotted as $\epsilon_V/\sqrt{3}$ in Fig.~\ref{plot_epscase2}) compared to its canonical picture, although at the cost of very steep rolling during inflation.

To compare how the predictions change with $n$, the same analysis has been done for $n=3$, where $c_s=1/\sqrt{5}$. Similar to the previous case, here too the power depends on a combination of model parameters: $\frac{P_{s}^{n=3}}{P_{s}^{n=1}}\propto \bigg(\frac{K_4\alpha ^2 \Lambda ^8}{f^4}\bigg)^{\frac{1}{5}}$. The inflation dynamics has been studied for this case as well for different values of $\beta/\alpha$.

This steep rolling eventually affects the graceful exit of inflation such that it becomes hard to obtain $50$-$60$ inflationary e-folds. Here, to achieve graceful exit with enough number of e-folds for both $n=2$ and $n=3$ cases, the respective combinations $(K_3\alpha\Lambda^4)^{1/3}$ and $(K_4\alpha^2\Lambda^8)^{1/5}$ were increased to high values for all the analyses with different $\beta$ values (see Fig.~\ref{plot3} for $n=2$).

Moreover, in such kinetic inflation cases with $c_s\neq 1$, the primordial non-Gaussianity is interesting to analyze as well. The non-Gaussianity parameter $f_{\rm NL}$ (equilateral) here can be expressed as~\cite{nonGauss}:
\begin{equation}
f_{\rm NL} \simeq -0.28u+0.02\frac{\epsilon}{\rho}s-1.53\epsilon-0.42\eta,
\end{equation}
where $u = 1-1/c_s^2$, $s=\dot{c_s}/(Hc_s)$; $\epsilon$ and $\eta$ are the Hubble slow-roll parameters. Therefore, for $n=2$, using Eqs.~\eqref{ns_kine} and~\eqref{rkine} with $c_s=1/\sqrt{3}$, 
\begin{equation}
f_{\rm NL}\simeq 0.87 - 0.3r - 0.31n_s.\label{kineNG}
\end{equation}
Similarly, for $n=3$, the non-Gaussianity parameter with $c_s=1/\sqrt{5}$ is:
\begin{equation}
f_{\rm NL}\simeq 1.47 - 0.468r - 0.35n_s.\label{kineNGn3}
\end{equation}
Therefore, once we obtain the inflationary observables $n_s$ and $r$, $f_{\rm NL}$ can be determined for different analyses.

\section{Results for the inflationary observables}
\label{results}
\setcounter{equation}{0}
\setcounter{figure}{0}
\setcounter{table}{0}
\setcounter{footnote}{1}
\begin{figure}[!htbp]
\centering{ \includegraphics[width=0.99\textwidth, height=0.6\textwidth]{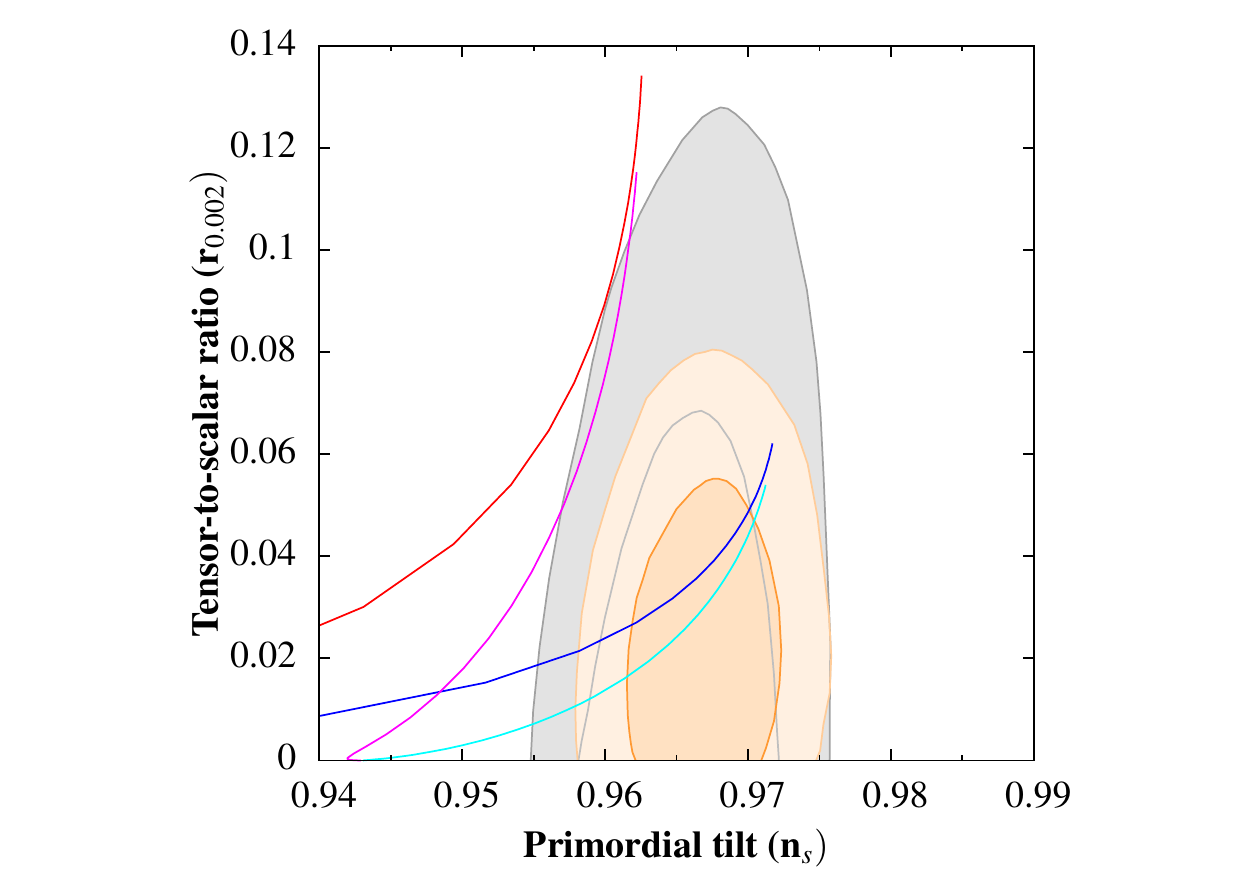} }
 \caption{\small{Comparison in the $n_s$-$r$ plane between canonical natural inflation (red), noncanonical natural inflation (blue), canonical Goldstone inflation (magenta) and noncanonical Goldstone inflation (cyan). The Goldstone inflation curves plotted here are for $\beta = 0.5$. The dark and light grey regions signify $68\%$ and $96\%$ confidence limits respectively for Planck TT,TE,EE+lowE+lensing data (2018)\cite{planck18}, whereas dark and light yellow regions signify $68\%$ and $96\%$ confidence limits respectively for Planck TT,TE,EE+lowE (2018)+lensing+BK14\cite{BICEP2}+BAO data \cite{BAO}.}}
\label{plot2}
\end{figure}
\begin{figure}[!htbp]
{\centering \includegraphics[width=0.8\textwidth, height=0.7\textwidth]{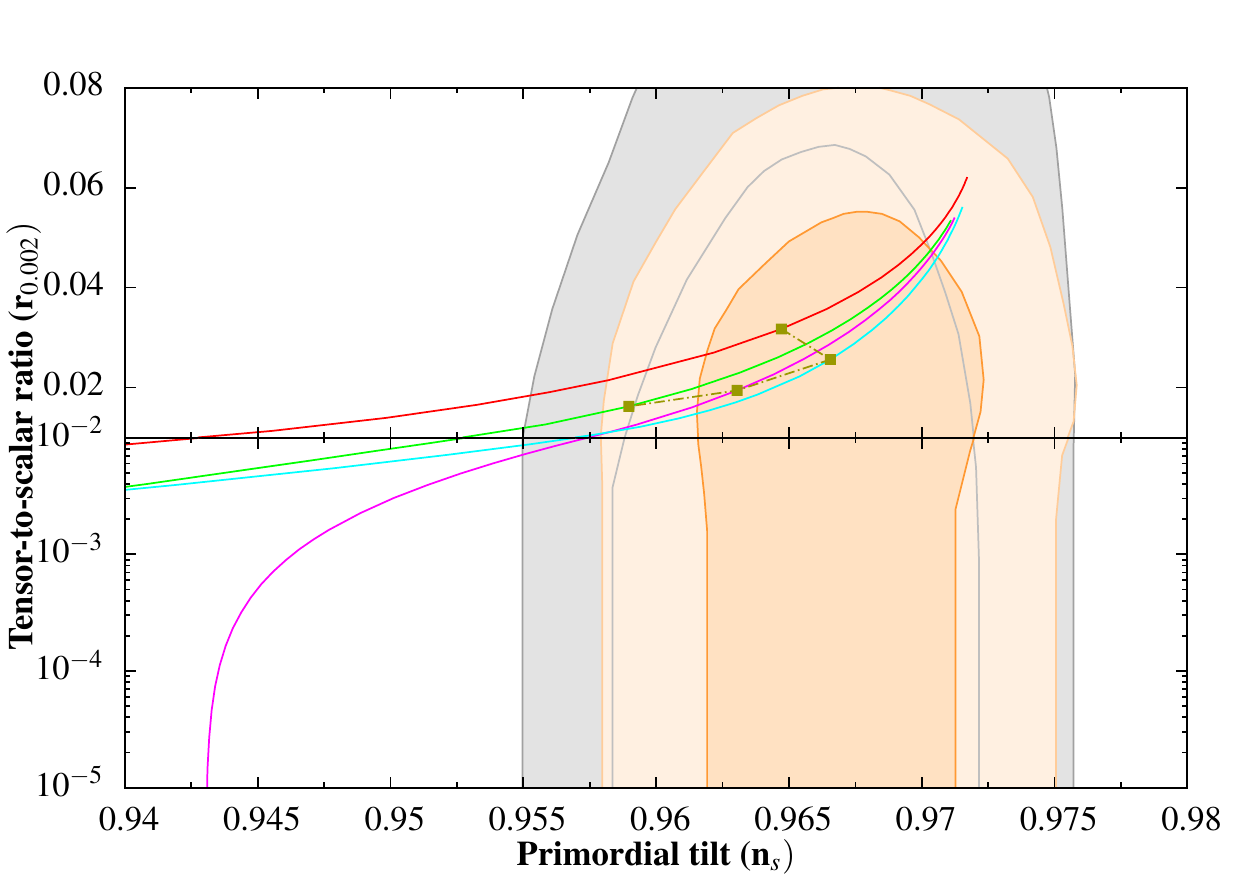} }
 \caption{\small{Comparison in the $n_s$-$r$ plane for noncanonical Goldstone inflation with different values of $\beta$ keeping $\alpha = 1$. $\beta = 0$ (noncanonical natural inflation) is in red, $\beta = 0.25$ in cyan, $\beta = 0.5$ in magenta and $\beta = 0.75$ in green. The green dot-dashed line connects the points with $f = 5M_{Pl}$ in all the curves. The grey dark and light regions signify $68\%$ and $96\%$ confidence limits respectively for Planck TT,TE,EE+lowE+lensing data (2018)\cite{planck18}, whereas yellow dark and light regions signify $68\%$ and $96\%$ confidence limits respectively for Planck TT,TE,EE+lowE (2018)+lensing+BK14 \cite{BICEP2}+BAO data\cite{BAO}.}}
\label{plot1}
\end{figure}
The main observables for inflation in CMB for the $\Lambda$CDM model are the scalar spectral index $n_s$ and the tensor-to-scalar ratio $r$ which are measured by Planck 2018\cite{planck18} with immense precision. The exact values of these parameters with $1\sigma$ errors as constrained by Planck 2018 are $n_s=0.9665\pm 0.0038$ (TT, TE, EE+ lowE+ lensing data ), $r<0.064$ (TT, TE, EE+ lowE +lensing data+ BK14). In this section, we discuss the predictions of the Goldstone inflation in the canonical regime for $n_s$ and $r$ with respect to their values in $1\sigma$ and $2\sigma$ confidence levels given by Planck 2018. We consider two different datasets in our analysis: (i) the most constrained Planck TT,TE,EE+lowE + lensing + BK14 + BAO and (ii) Planck TT,TE,EE+lowE+lensing.

In Fig.~\ref{plot2}, we compared the predictions for natural inflation and Goldstone inflation in the canonical regime and in the noncanonical regime ({\it Case 1}). The noncanonical plot here is just for comparison and plotted for $\beta = 0.5$ ($C_{\Lambda}=1$, $\alpha=1$). It is evident from this plot that noncanonical picture $K_{nc}(\phi)=V(\phi)/\Lambda^4$ does improve the predictions for inflation by a significant suppression of $r$. 

In Fig.~\ref{plot1}, we explored the observables in the $n_s$-$r$ plane for {\it Case 1} of noncanonical Goldstone inflation while varying the model parameter $\beta$. For each value of $\beta$, the solid line runs for variation of the breaking scale $f$ up to $16M_{Pl}$. The plot shows that for most of the super-Planckian breaking scales $f>M_{Pl}$, the noncanonical scenario ({\it Case 1}) lowers the tensor-to-scalar ratio $r$ for all values of $\beta < 1$. But, similar to the default canonical Natural inflation case, this does not improve the predictions for $n_s$ and $r$ in the sub-Planckian scales $f<M_{Pl}$. This was hinted from Fig.~\ref{plot_epscase1}, where the pivot field value for the noncanonical {\it Case 1} predicted higher value of $\epsilon_{V}$ for $f\leq M_{Pl}$. Particularly, for $\beta = 0.5$, even though $r$ decreases rapidly with the decrease in $f$ below $M_{Pl}$, the spectral index $n_s$ is outside the current precision bounds by Planck. 
\begin{figure}[!htbp]
  \centering
  \subfloat[]{\includegraphics[width=0.5\textwidth]{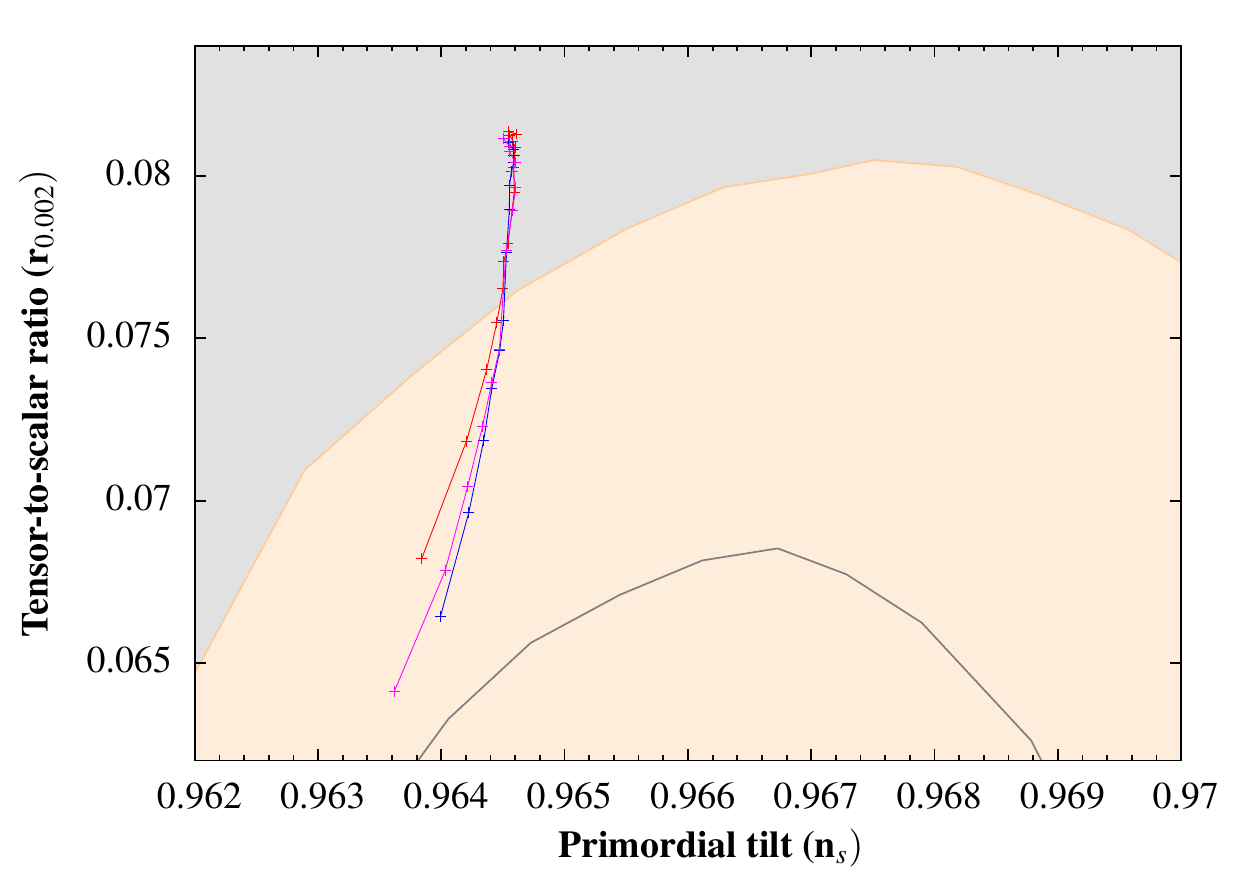}\label{plot3}}
  \hfill
  \subfloat[]{\includegraphics[width=0.5\textwidth]{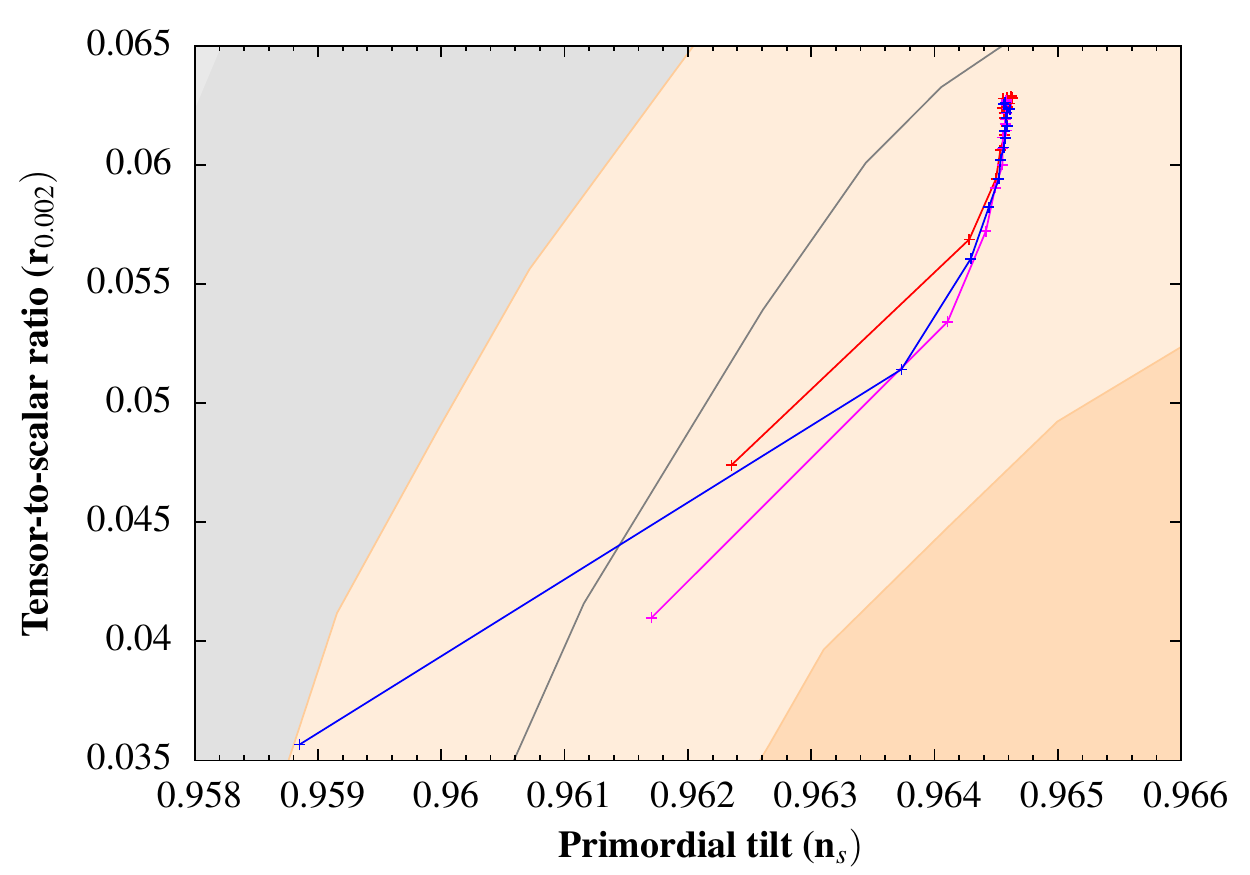}\label{plot3_n3}}
  \caption{\small{The $n_s$-$r$ plot for kinetic inflation for $n=2$ in the left panel and $n=3$ in the right panel. For both cases here, kinetic natural inflation curve is plotted in red, whereas kinetic Goldstone inflation curves for $\beta = 0.2$ is in magenta, for $\beta = 0.5$ is in blue. The light grey region pervading all through the plot signifies $96\%$ confidence limit for Planck TT,TE,EE+lowE+lensing data (2018)\cite{planck18}, whereas the dark grey contour signifies $68\%$ confidence limit for the same data combination. The yellow shaded region signifies $96\%$ confidence level for Planck TT,TE,EE+lowE (2018)+lensing+BK14\cite{BICEP2}+BAO data\cite{BAO}. For each curve, the lowest value of $r$ is for $f=0.5M_{Pl}$.}}
  \label{kinensr2plots}
\end{figure}
\begin{figure}[htbp]
  \centering
  \subfloat[]{\includegraphics[width=0.5\textwidth]{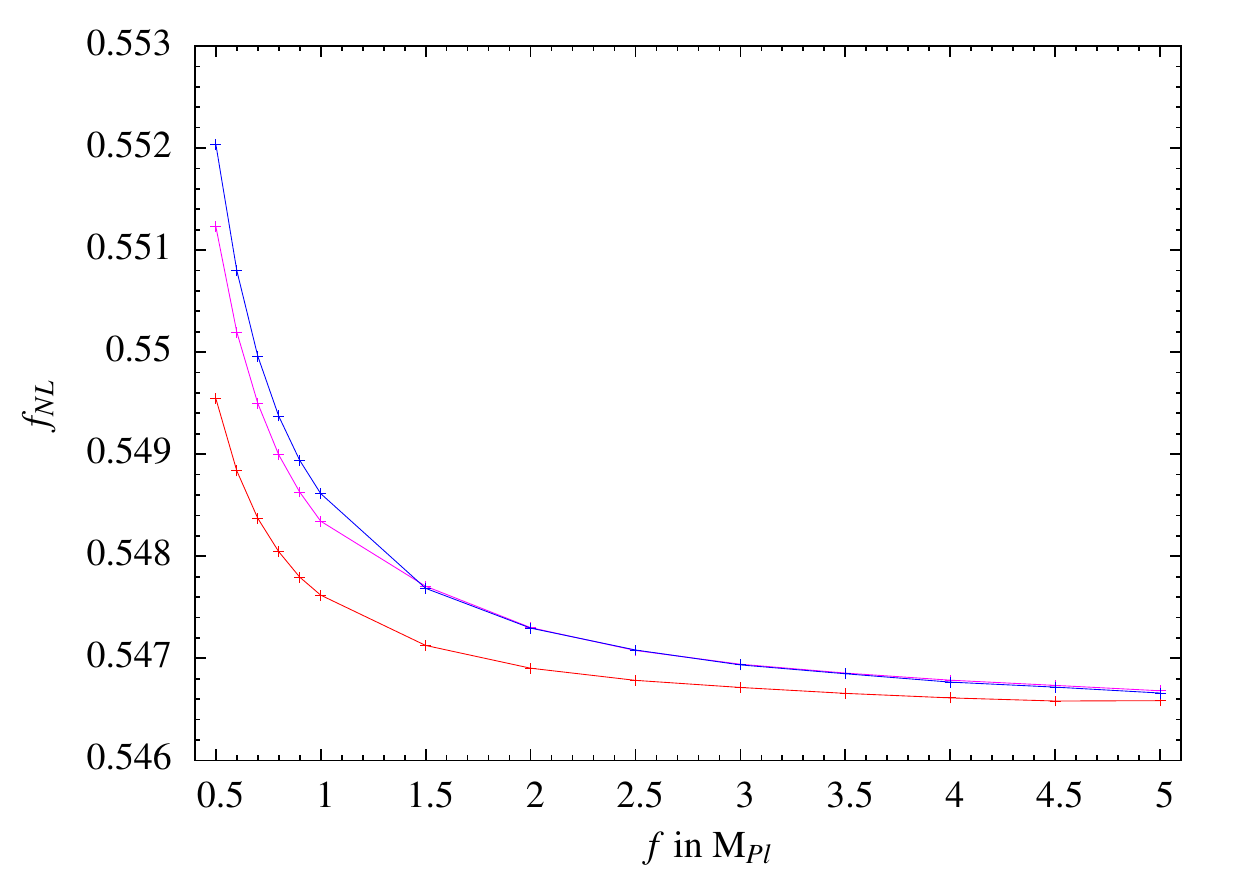}\label{plot_NG}}
  \hfill
  \subfloat[]{\includegraphics[width=0.5\textwidth]{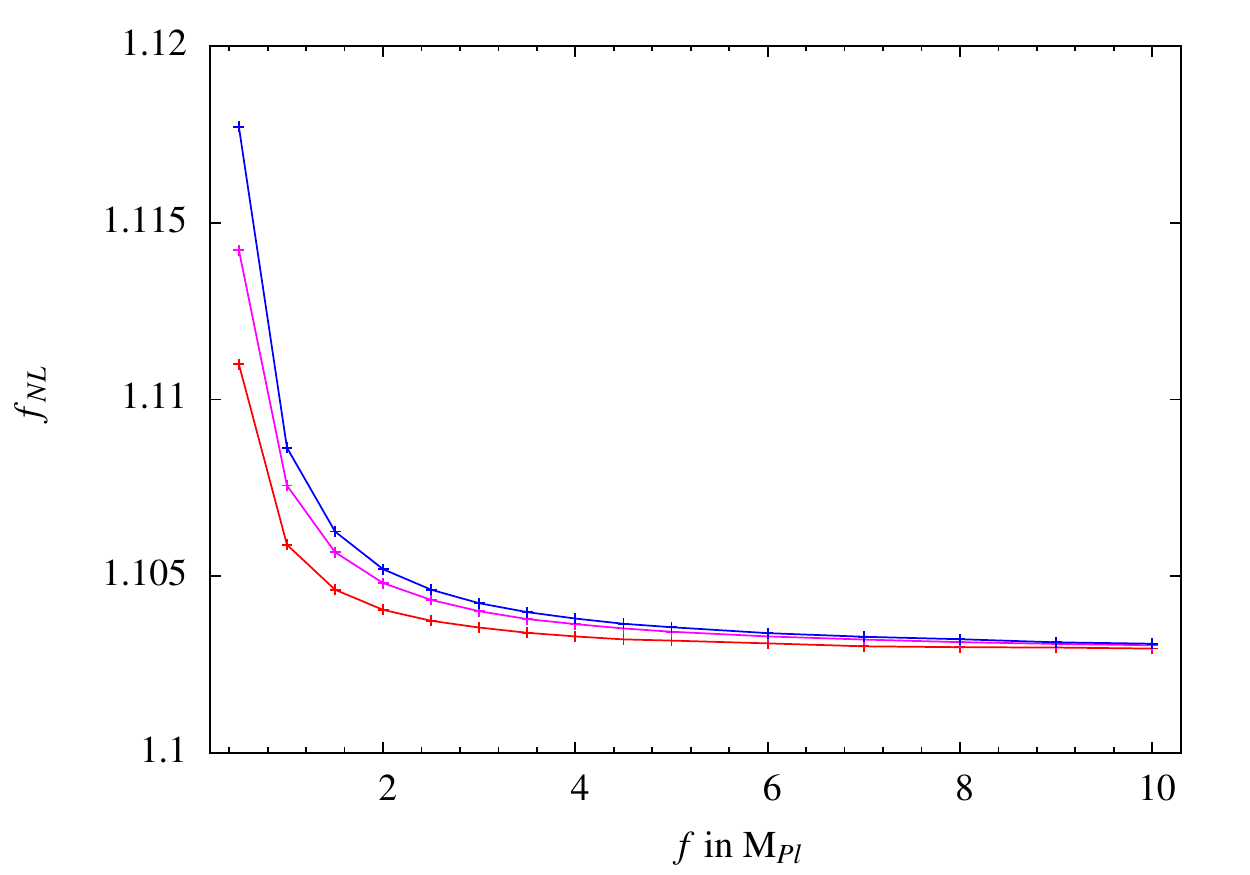}\label{plotNG_n3}}
  \caption{\small{For kinetic inflation, non-Gaussianity parameter $f_{\rm NL}$ varied with $f$ where $\beta=0$ (natural inflation) is plotted in red, $\beta = 0.2$ in magenta and $\beta = 0.5$ in blue. $n=2$ and $n=3$ are plotted in the left and right panels respectively.}}
  \label{NGplots}
\end{figure}
\begin{figure}[!htbp]
{\centering \includegraphics[width=0.9\textwidth, height=0.6\textwidth]{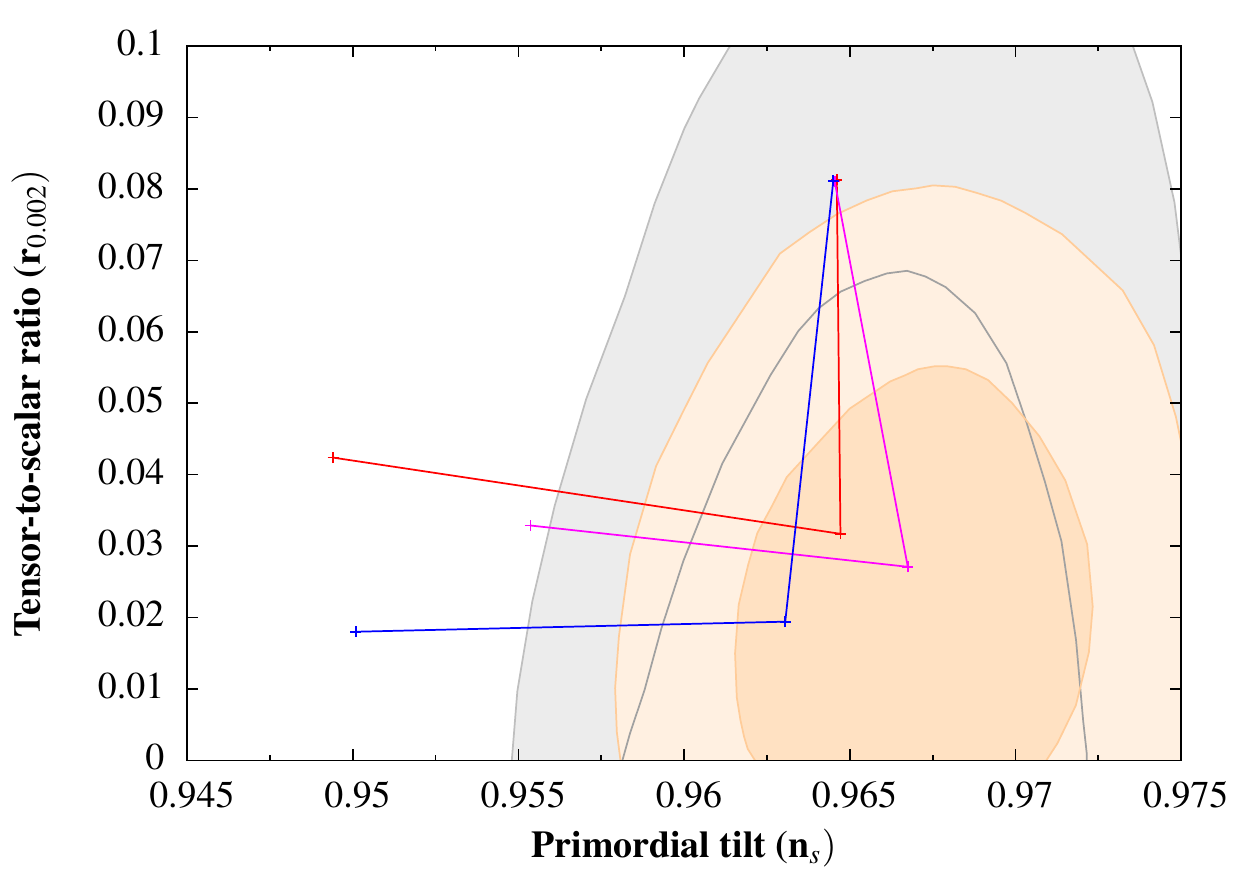} }
 \caption{\small{Comparison in the $n_s$-$r$ plane between natural inflation and Goldstone inflation for $f=5M_{Pl}$ (fixed). The natural inflation curve is plotted in red whereas Goldstone inflation curve for the combination $\alpha = 1$,$\beta = 0.5$ is in blue and for the combination $\alpha = 1$,$\beta = 0.2$ is in magenta. The three points from left to right in each of the curves are for the canonical, noncanonical and kinetic cases respectively. The dark and light grey regions signify $68\%$ and $96\%$ confidence limits respectively for Planck TT,TE,EE+lowE+lensing data (2018)\cite{planck18}, whereas dark and light yellow regions signify $68\%$ and $96\%$ confidence limits respectively for Planck TT,TE,EE+lowE (2018)+lensing+BK14\cite{BICEP2}+BAO data\cite{BAO}.}}
\label{plot4}
\end{figure}

Fig.~\ref{kinensr2plots} shows the predictions for the inflationary observables for {\it Case 2} of noncanonical Goldstone inflation for $n=2$ (Fig.~\ref{plot3}) and $n=3$ (Fig.~\ref{plot3_n3}). Here, we see that the $n_s$ and $r$ values for the sub-Planckian breaking scales $f<M_{Pl}$ are inside the $2\sigma$ bounds give by Planck dataset (i) for all values of $\beta$. But the prediction of $r$ is larger compared to {\it Case 1}, which makes the Goldstone inflation in kinetic noncanonical regime {\it Case 2} vulnerable to future precision detections of primordial tensor modes. 

Fig.~\ref{NGplots} shows variation of $f_{\rm NL}$ with $f$ for different $\beta$ for $n=2$ in Fig.~\ref{plot_NG} using Eq.~\ref{kineNG} and for $n=3$ in Fig.~\ref{plotNG_n3} using Eq.~\ref{kineNGn3}. For all the cases we find $f_{\rm NL} \sim \mathcal{O}(1)$. The recent observations refer to $f_{\rm NL} = -4 \pm 43$~\cite{Renaux-Petel:2015bja} and therefore, in this case it is well within the allowed value. Though the estimate on the non-Gaussianity  will improve dramatically in upcoming observations such as Cosmic Origin Explorer (COrE)\cite{finelli}, and then models like this, which predict very small but positive $f_{\rm NL}$, can be tested. The small value of $f_{\rm NL}$ for $n=2$ can be attributed to it being the minimal deviation from the canonical case. The $n=3$ case already shows increase in $f_{\rm NL}$ and higher orders in $X$ can increase $f_{\rm NL}$ further. But, in this work we limit our analysis to $n=2$ and $n=3$ because, first, renormalization of the theory is an issue in any case of kinetic inflation and higher $n$ are riskier to deal with. Second, the observational bound on the cosmological sound speed $c_s$ restricts the power $n$ of the kinetic term. The $2\sigma$ upper bound on $f_{\rm NL}^{Equil}<82$ provides lower bound for $c_s=\frac{1}{\sqrt{2n-1}}$~\cite{Tzirakis:2008qy}. The exact bound on $c_s$ would depend on the exact kinetic form.

 

In Fig.~\ref{plot4}, the three solid lines all refer to the same $f$, but differ in inputs of $\beta$. In each of the solid lines, we have compared the three cases: canonical (leftmost point), noncanonical {\it Case 1} (middle point) and noncanonical {\it Case 2} ($n=2$) (rightmost point). As hinted in the previous figures, we can see that of the all three cases, the noncanonical {\it Case 1} provides best predictions for the super-Planckian case $f=5M_{Pl}$ with reference to current bounds from Planck.
\section{Reheating}
\label{reheat}
\begin{figure}[!htb]
    \centering
    \includegraphics[width=12.0cm]{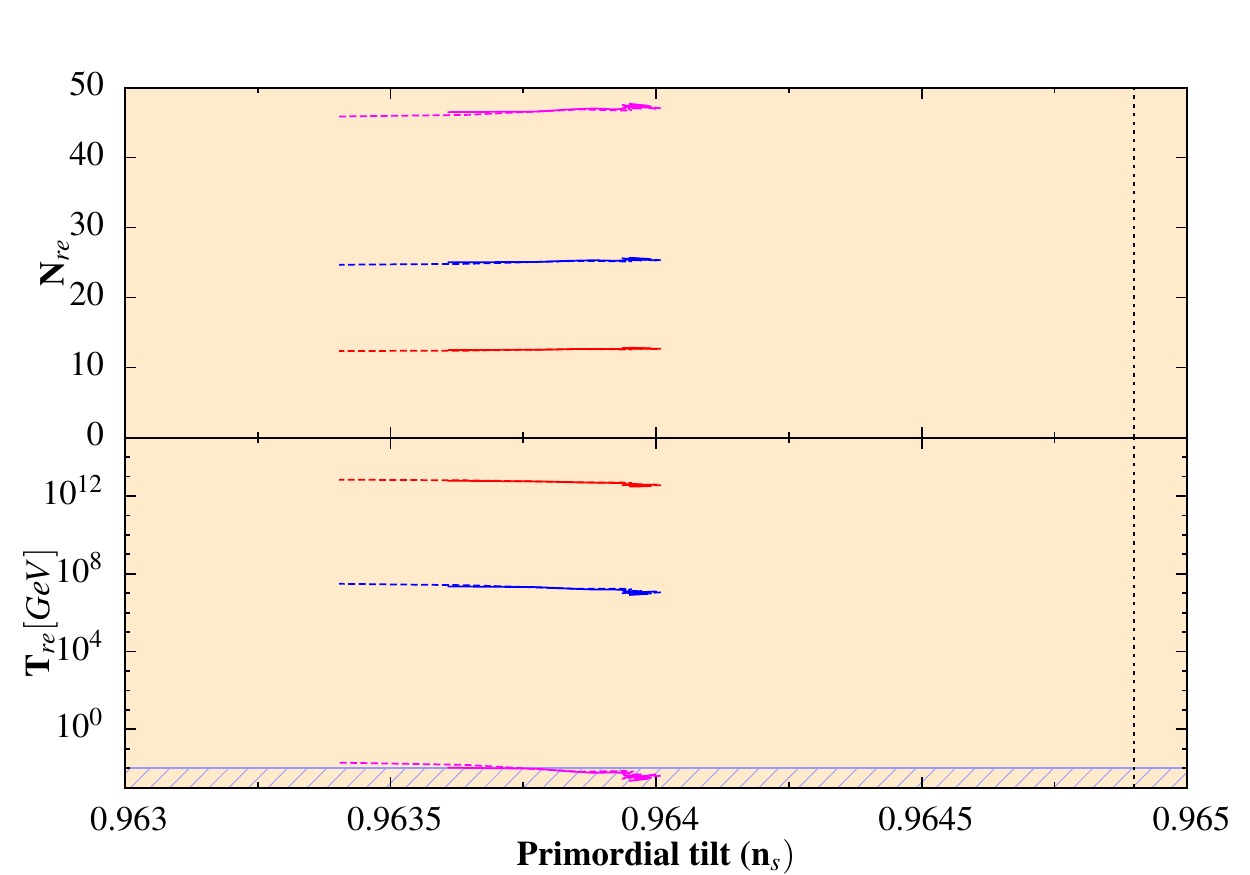} 
             \caption{\small The red ($w_{\rm re}= -1/3$), blue ($w_{\rm re}=0$), magenta ($w_{\rm re} = 2/13$) lines represents different equation of states for $\beta/\alpha=0$ (solid lines) and $\beta/\alpha=0.2$ (dashed lines). For both $\beta/\alpha$ values, increasing $w_{\rm re}$ beyond 2/13 gives $T_{\rm re}<T_{\rm BBN}$, even for $f=0.5 M_{\rm Pl}$. The brown region pervading all through the background of the plot signifies that the plotted $n_s$ values are inside $1\sigma$ for Planck TT,TE,EE+lowE+lensing+BK14+BAO. The black dashed line marks the central value for the same data combination, which provides the most stringent CMB bounds on $n_s$. The blue mesh region in $T_{\rm re}$ plot signifies $T_{\rm re}<T_{\rm BBN}$.} 
\label{fig: reh}
\end{figure} 

After the end of inflation, an era of reheating is required to have the Universe ready to enter the radiation domination and follow it with a big bang nucleosynthesis (BBN). Although, the era of reheating is still not very well understood but one can probe the era by parameterizing in terms of reheating temperature ($T_{\rm re}$), number of e-folds during reheating ($N_{\rm re}$) and effective equation of states during reheating ($w_{\rm re}$). The mechanism of reheating to the Standard Model (SM) degrees of freedoms can be a difficult task to implement as elaborately discussed in the following papers \cite{mazumdar1} and \cite{mazumdar2}. We want to emphasize that the objective of this analysis is to get the idea about the reheating parameters without going in details about the particular mechanism. Reheating being a very complex phase to analyze, this analysis gives us an approximate idea of what to expect.

If $w_{\rm re}$ remains constant during reheating then energy density can be related to the scale factor $a$ (using $\rho_{\rm re}\propto a^{-3(1+ w_{\rm re})}$) as:
\begin{equation}
\frac{\rho_{\rm end}}{\rho_{\rm re}}= \bigg(\frac{a_{\rm end}}{a_{\rm re}}\bigg)^{-3(1+ w_{\rm re})}
\end{equation}
As usual, the subscripts ``end'' and ``re'' imply the values of the quantities at the end of inflation and reheating respectively. Now following \cite{naut} one can write:
\begin{eqnarray}
N_{\rm re}&=& \frac{4}{1-3w_{\rm re}}\bigg(-\frac{1}{4}\ln\bigg(\frac{3^2\times5}{\pi^2g_{\rm re}}\bigg)-\frac{1}{3}\ln\bigg(\frac{11g_{\rm re}}{43}\bigg)-\ln\bigg(\frac{V_{\rm end}^{1/4}}{H_k}\bigg)-\ln\bigg(\frac{k}{T_0a_0}\bigg)-N_k\bigg),\label{Nre_k}\\
T_{\rm re}&=& \bigg(\bigg(\frac{43}{11g_{\rm re}}\bigg)^{1/3}\frac{a_0T_0}{k}H_ke^{-N_k}\bigg(\frac{3^2\times5\times V_{end}}{\pi^2g_{\rm re}}\bigg)^{\frac{1}{3(1+w_{\rm re})}}\bigg)^{\frac{3(1+w_{\rm re})}{3w_{\rm re}-1}}.\label{Tre_k}
\end{eqnarray}
Here, $g_{\rm re}$ the number of degrees of freedom during reheating, which is roughly taken to be $10^2$, $a_0=1$ is the scale factor today. The current CMB temperature is $T_0$. Now, in our analysis, we have fixed the number of e-folds at the horizon exit $N_k$ dynamically at $55$ where $k$ is taken to be $0.002{\rm Mpc^{-1}}$. Here, it is evident from Eq.~\eqref{Nre_k} that the reheating dynamics can only be solved when $w_{\rm re}\neq 1/3$. It would be interesting to check the effects of non-Gaussianity from the (p)reheating phase as discussed in\cite{mazumdarng1, mazumdarng2}. In noncanonical scenarios, the phase of (p)reheating might be unavoidable. It would be interesting to check the effects and we intend to come up with this analysis elsewhere in future.

The main result of reheating for noncanonical Goldstone inflation {\it Case 2} ($n=2$) is depicted in Fig.~\ref{fig: reh}. This figure clearly shows that going beyond $w_{\rm re} = 2/13$ does not allow reheating to end above the BBN temperature $T_{\rm BBN} \sim 10 {\rm MeV}$, even for $f=0.5M_{\rm Pl}$. This is exactly what is expected for the kinetic inflation as here, equation of state follows that of the stiff fluid. Thus, without the detailed physical knowledge of the reheating era one can still realize whether at the end of kinetic inflation one can get a proper reheating era, which is clearly achieved here for $w_{re} \leq 2/13$. This practice of probing reheating via $N_{\rm re}$, $T_{\rm re}$ and $w_{\rm re}$ without the proper knowledge of the dynamics of the era, has been carried out in literature~\cite{kumar, sukmaha, kahler, copeland} for standard and alternate thermal history of our Universe.

\section{Conclusions and Discussions}
\label{discussions}
\setcounter{equation}{0}
\setcounter{figure}{0}
\setcounter{table}{0}
\setcounter{footnote}{1}
With future observations like {\it CMB-S4} \cite{abzajian} and {\it COrE} \cite{finelli} with promising prospects to measure the spectral tilt very precisely ($\Delta n_s \sim 0.002$), and with future possibilities to constrain the primordial tensor modes, a systematic study of the unconventional scenarios of inflation for theoretically motivated models has become essential. Models that are well motivated from theory but facing trouble to predict observable parameters within experimental bounds need to be reevaluated in scenarios such as nonminimal coupling to gravity~\cite{nozari} or noncanonical inflation. Inflaton being a pNGB has a very promising theoretical justification and therefore, a Goldstone potential to drive the inflationary expansion is studied here in the noncanonical scenario constrained from latest CMB data.

We emphasize that using a noncanonical framework in this work helped to avoid fine tuning of model parameters, which is unavoidable in the canonical case of Goldstone inflation. For {\it Case 1}, the prototype $K_{nc}(\phi) = V({\phi})/\Lambda^4$ is just to give an effective flatness to the potential. More forms of $K_{nc}(\phi)$ arising from nonminimal gravitational coupling will be interesting to analyze, as they come naturally from nontrivial Lagrangians in the Jordan frame \cite{Broy:2015qna,Pallis:2010wt,sb}. 

For noncanonical {\it Case 1} we get smaller tensor-to-scalar ratio ($r$), however we do not achieve enough e-folds of inflation for sub-Planckian $f$. On the other hand, for {\it Case 2} we achieve $\sim 55$ e-folds of inflation even for sub-Planckian $f$, but at the cost of $r$ values lying outside the current 68$\%$ bound. A generalized kinetic term with both the cases switched on will be interesting in terms of the prediction for observables, if their effects combine in a constructive manner. The next natural step should be to test these models with thorough numerical analysis using Bayesian techniques. Another exciting case would be to check the effects of noncanonical inflation in the braneworld scenarios. As expected in braneworld scenario, there is a natural tendency of increasing $r$ \cite{mrg}, it would be interesting to check NCI in that paradigm. We hope to return to these problems in near future. Another issue which might need a serious theoretical explanation is the observed anomaly at the low multipole in the CMB power spectrum as observed by Planck as well as WMAP. Many explanations \cite{lowl1}-\cite{lowl5} are being put forward and on that note it would be exciting to check if a noncanonical initial condition could orchestrate such an imprint on such scales. 

Although we approach {\it Case 2} in this paper in a phenomenological sense with \textit{ad hoc} introduction of higher order derivative terms in the inflaton Lagrangian, there are very interesting theoretical implications of such terms. General theoretical considerations of an inflaton Lagrangian ignores higher order derivative terms for two major reasons: (i) assuming that the initial kinetic energy is very small in the initial Hubble patch and a slow roll ensures that it stays this way~\cite{Linde:2005ht,Linde:1993xx}; (ii) even if one starts with a large initial field velocity, attractor solution can drive inflaton to a slow roll regime very rapidly\cite{Linde:1993xx,mukhanov2005}. However, if one considers the full Lagrangian with every possible higher order derivative term (e.g. $f(\Box)$ is a polynomial in the $\Box$ operator~\cite{picon,mukhanov,Chen:2006nt,Burrage:2010cu}) then it leads to additional ghostlike states if we aim toward a nontachyonic theory. These ghostlike states make the vacuum unstable and the corresponding propagator contains negative residues~\cite{Biswas:2005qr}. Therefore, to have viable ghostfree higher derivative kinetic terms in the inflaton Lagrangian, very specific functional forms of $f(\Box)$ are allowed, which is discussed, e.g., in case of $p$-adic string inflation theories~\cite{Barnaby:2006hi,Freund:1987ck,Freund:1987kt,Brekke:1988dg}. Although implementing a theoretically viable full kinetic term containing higher order derivatives here can lead to a more complete description of kinetic Goldstone inflation, this paper introduces only the first possible correction as {\it Case 2}.

Finally, we comment regarding the recently proposed swampland criteria (fiasco)~\cite{vafa} which created some sensation in the cosmology community. On that regard, we would like to emphasize that a noncanonical inflation, specifically {\it Case 2}, with a theoretically well-motivated potential could actually evade the problem and might be a natural answer to it since the Lagrangian for NCI is expected and motivated from string theory. The bounds on $c_s$ from CMB could also play a key role in that as indicated in~\cite{kinney}. This is another interesting problem that we would like to address soon. \\
\textbf {Acknowledgments:} S.B. is supported by institute postdoctoral fellowship from Physical Research Laboratory. MRG and SB want to thank G.~Mathews, K.~Dutta, A.~Banerjee and N.~Kumar for useful discussions. SB sincerely thanks A.~Nautiyal for his comments and discussions in the early phases of this work. Work of MRG is supported by Department of Science and Technology, Government of India under the Grant Agreement number IF18-PH-228 (INSPIRE Faculty Award). SB and MRG sincerely thank the referee for suggesting the reheating analysis and relevant literature survey regarding this work.


\end{document}